\documentclass[aps, prl,amsmath,amssymb, reprint, superscriptaddress]{revtex4-2}

\usepackage{graphicx}
\usepackage{dcolumn}
\usepackage{bm}
\usepackage{xcolor}
\usepackage[T1]{fontenc}

\begin{document}

\title{Mutual synchronization of two \\asymmetric-nano-constriction-based spin-Hall nano-oscillators}

\author{Roman V. Ovcharov}
\affiliation{ 
Department of Physics, University of Gothenburg, Gothenburg 41296, Sweden
}

\author{Roman S.  Khymyn}
\affiliation{ 
Department of Physics, University of Gothenburg, Gothenburg 41296, Sweden
}

\author{Akash Kumar}
\affiliation{ 
Department of Physics, University of Gothenburg, Gothenburg 41296, Sweden
}
\affiliation{
Center for Science and Innovation in Spintronics, Tohoku University, Sendai 980-8577, Japan
}
\affiliation{
Research Institute of Electrical Communication, Tohoku University, Sendai 980-8577, Japan
}

\author{Johan \AA kerman}
\affiliation{ 
Department of Physics, University of Gothenburg, Gothenburg 41296, Sweden
}
\affiliation{
Center for Science and Innovation in Spintronics, Tohoku University, Sendai 980-8577, Japan
}
\affiliation{
Research Institute of Electrical Communication, Tohoku University, Sendai 980-8577, Japan
}

\date{\today}

\begin{abstract}
We propose an asymmetric-nanoconstriction (ANC) design of spin-Hall nano-oscillators (SHNOs) and investigate mutual synchronization of a pair of such devices using micromagnetic simulations. The ANC geometry enables strong dipolar coupling at sub-50 nm separations while preserving independent current bias for each oscillator. We first characterize the auto-oscillation of a single ANC-SHNO, revealing a broad frequency tuning range and a field-controlled crossover between negative and positive nonlinearities. We then demonstrate that two such oscillators can mutually synchronize solely via dipolar stray fields, without electrical or spin-wave coupling. Depending on the bias conditions, the coupled pair exhibits robust in-phase (0$^\circ$) or out-of-phase (180$^\circ$) locking. Notably, we find a bias-dependent amplitude correlation: when the oscillators sustain comparable amplitudes, both in-phase and out-of-phase synchronization are accessible, whereas amplitude imbalance drives the system into an out-of-phase state accompanied by suppression of the weaker oscillator. By combining strong conservative coupling with independent frequency and gain control, the ANC-SHNO platform provides a scalable route toward phased oscillator arrays, neuromorphic computing architectures, and experimental exploration of non-Hermitian spintronic dynamics.
\end{abstract}

\maketitle

Spintronic nano-oscillators (SNOs) are an emerging class of nanoscale microwave signal generators driven by spin-orbit or spin-transfer torques in magnetic thin films~\cite{chen2016spin,kumar2024mutual}. These oscillators exhibit rich nonlinear dynamics, and networks of them can serve as model systems for studying synchronization phenomena at the nanoscale \cite{castro2022mutual,erokhin2014robust,kendziorczyk2016mutual,awad2017long}. 
A long-standing motivation for coupling multiple SNOs is to increase their combined microwave power and improve signal coherence by reducing the linewidth~\cite{kaka2005mutual, mancoff2005phase,houshang2016spin}. Beyond microwave generation, SNO networks are being explored as hardware building blocks for in‑memory and neuromorphic computing~\cite{chumak2022advances, finocchio2024roadmap,gonzalez2024spintronic, arunima2025controllable, manna2024phase}. For example, coupled networks of SNOs have been mapped onto the Ising spin model for combinatorial optimization~\cite{albertsson2021ultrafast, houshang2022phase}, and neuromorphic-computing schemes have utilized the nonlinearity of SNO dynamics for signal classification and pattern recognition~\cite{romera2018vowel, zahedinejad2020two, romera2022binding, zahedinejad2022memristive}. Therefore, developing reliable, tunable synchronization in dense SNO arrays serves both RF engineering and emerging spintronic computing concepts.

Spin‑Hall nano‑oscillators (SHNOs) are an emerging subclass of SNOs in which auto‑oscillations are sustained by a dc current via the spin Hall effect in a heavy‑metal/ferromagnet bilayer, rather than by a spin‑polarized current from a fixed magnetic layer as in spin‑torque nano‑oscillators (STNOs)~\cite{demidov2014nanoconstriction}. This key difference eases nanofabrication, thus enabling flexible layouts (with feature sizes down to 10~nm~\cite{behera2024ultra}), broader material choices, and direct access to individual oscillators (e.g., for optical probing or local gating)~\cite{hache2025nanoscale, muralidhar2022optothermal,kumar2022fabrication}. The most widely used implementation is the nano‑constriction SHNO (NC‑SHNO)~\cite{demidov2014nanoconstriction}, where a narrow constriction funnels current and localizes auto-oscillating modes~\cite{dvornik2018origin}. Mutual synchronization between two or more NC-SHNOs (up to 100,000) on a continuous bilayer structure has been achieved through combined spin-wave and dipolar mechanisms, as well as electrical coupling via common microwave currents~\cite{mazraati2022mutual, behera2025ultra}. Thus, SHNO arrays offer a versatile platform to study the statistical mechanics of complex oscillator networks, including scale-free architectures. However, when coupling is mediated by propagating spin waves~\cite{kumar2025spin, haidar2024interference}, the finite group velocity introduces distance- and frequency-dependent phase lags in coupling, which makes it difficult to achieve global phase locking and limits robust synchronization~\cite{yeung1999time, earl2003synchronization}. A practical route to avoid this delay problem is to favor (quasi-)instantaneous near-field, i.e., pure dipolar, coupling, for example, by breaking magnetic continuity between nodes.

\begin{figure}[hbt!]
    \centering
    \includegraphics[width=\linewidth]{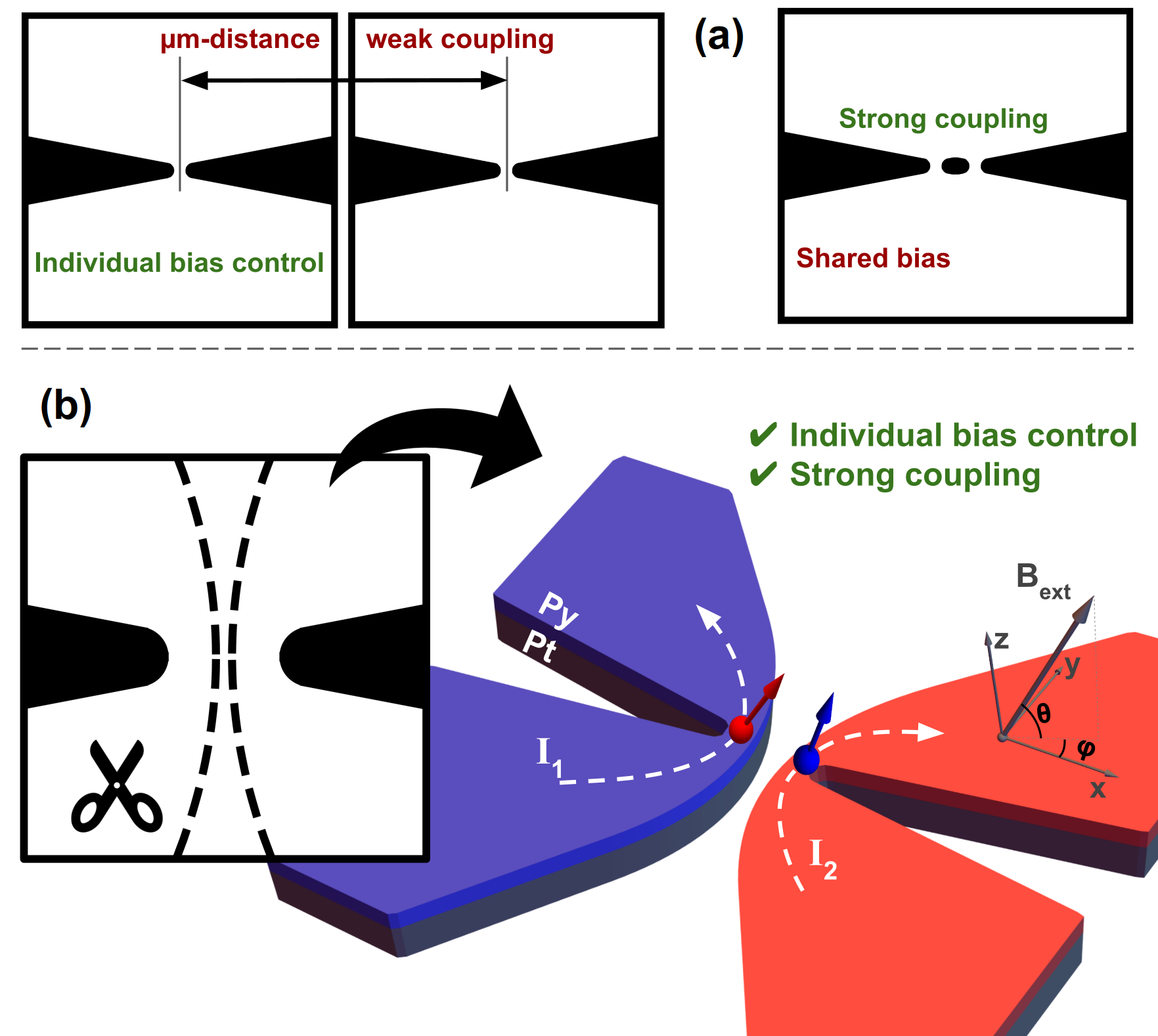}
    \caption{a) Trade-offs between independent bias and strong coupling in conventional bowtie-shaped NC SHNOs. b) Geometry of the proposed asymmetric-NC SHNO pair. A dc current $I_1$ ($I_2$) driven through the heavy‑metal/ferromagnet bilayer of the left (right) device generates a transverse spin current that excites a localized auto‑oscillation at the bent edge of each constriction. The edges face one another across a 30~nm gap, so that the oscillators couple exclusively through their magneto‑dipolar fields. Colored arrows depict the instantaneous magnetization at the oscillation centers. A static external field $B_\text{ext}$ is applied at an oblique angle to the film plane, as indicated by the inset coordinate axes.}
    \label{fig:schema}
\end{figure}

A purely magneto‑dipolar synchronization of independently biased SHNOs, however, has not yet been reported. In conventional (bowtie-shaped) NC-SHNOs, the constrictions are patterned on a micrometre-wide mesa, yielding centre-to-centre distances where stray-field coupling alone is too weak to enforce mutual synchronization~\cite{kumar2024mutual}. To close this gap, we introduce a new design for NC-SHNOs called the asymmetric‑nano‑constriction SHNO (ANC‑SHNO) shown in Fig.~\ref{fig:schema}. Conceptually, the design slices a classical constriction in half and gently bends its edges, producing a single‑edge oscillator with its peculiarities, such as asymmetry in the auto-oscillation mode profile. Crucially, two ANC-SHNOs can be placed at any edge-to-edge distance, bringing the dipolar interaction into the strong-coupling regime required for geometry-limited synchronization, while retaining independent current bias for each oscillator~\cite{slavin2009nonlinear}. 
In particular, the ability to tune each oscillator's current independently (and thus its frequency and effective gain) opens the door to exploring non-Hermitian oscillator dynamics, as recently demonstrated in vortex-based STNOs~\cite{wittrock2024non,perna2024coupling, matveev2023exceptional}. Non-Hermitian physics refers to open, non-conserving systems that exchange energy with their environment, often through an imbalance of gain and loss~\cite{hurst2022non}. Such systems can host exceptional points -- spectral degeneracies where both eigenfrequencies and eigenmodes coalesce -- leading to abrupt transitions in oscillation behavior. In spintronics, such conditions arise naturally from the competition between spin-torque–induced negative damping and intrinsic magnetic relaxation, and ANC-SHNOs with their independent gain–loss tuning provide a minimal and versatile platform for accessing these regimes.

\bigskip
We evaluate the proposed architecture with GPU‑accelerated micromagnetic simulations performed in MuMax3 \cite{vansteenkiste2014design}. The material parameters are chosen to represent a typical spin-Hall oscillator stack of a 5~nm‑thick Ni$_{80}$Fe$_{20}$ (Py) free layer on a 5~nm Pt underlayer. The following parameters are used: exchange stiffness $A_\text{ex} = 10$ pJ/m, saturation magnetization $M_s = 600$ kA/m, Gilbert damping $\alpha = 0.02$, gyromagnetic ratio $\gamma/2\pi = 29.53$ GHz/T. For comparison across different nonlinear regimes, we apply a magnetic field of $B_\text{ext}=1$~T at a $62^{\circ}$ out-of-plane angle. This configuration yields a weak analytically calculated magnon-magnon coupling, which can then evolve with the auto-oscillation mode~\cite{tm2025tunable, rajabali2023injection}, as will be shown below. Importantly, the weak nonlinearity provides a favorable balance for sustaining high-volume auto-oscillating modes while avoiding excessive spin-wave-emission losses~\cite{slonczewski1999excitation} or the reduced-volume penalties associated with overly localized (``bullet''-like) modes~\cite{slavin2005spin}.

The simulated device geometry consists of a rectangular ferromagnetic region with a thickness of 5~nm and a 60 nm-wide notch at one end to form the asymmetric constriction. For the two-oscillator simulations, two such regions are placed so that their notched edges face each other at 30~nm edge-to-edge separation. The heavy metal underlayers providing spin Hall current are included implicitly via a spatially non-uniform spin-torque term. In the two-oscillator configuration, the current distribution is split into two such regions, ensuring independent electrical control of each oscillator. From the steady-state magnetization time traces, we extract the power spectral density (PSD) to determine each oscillator's auto-
oscillation frequency, amplitude, and phase.

\begin{figure}[hbt!]
    \centering
    \includegraphics[width=\linewidth]{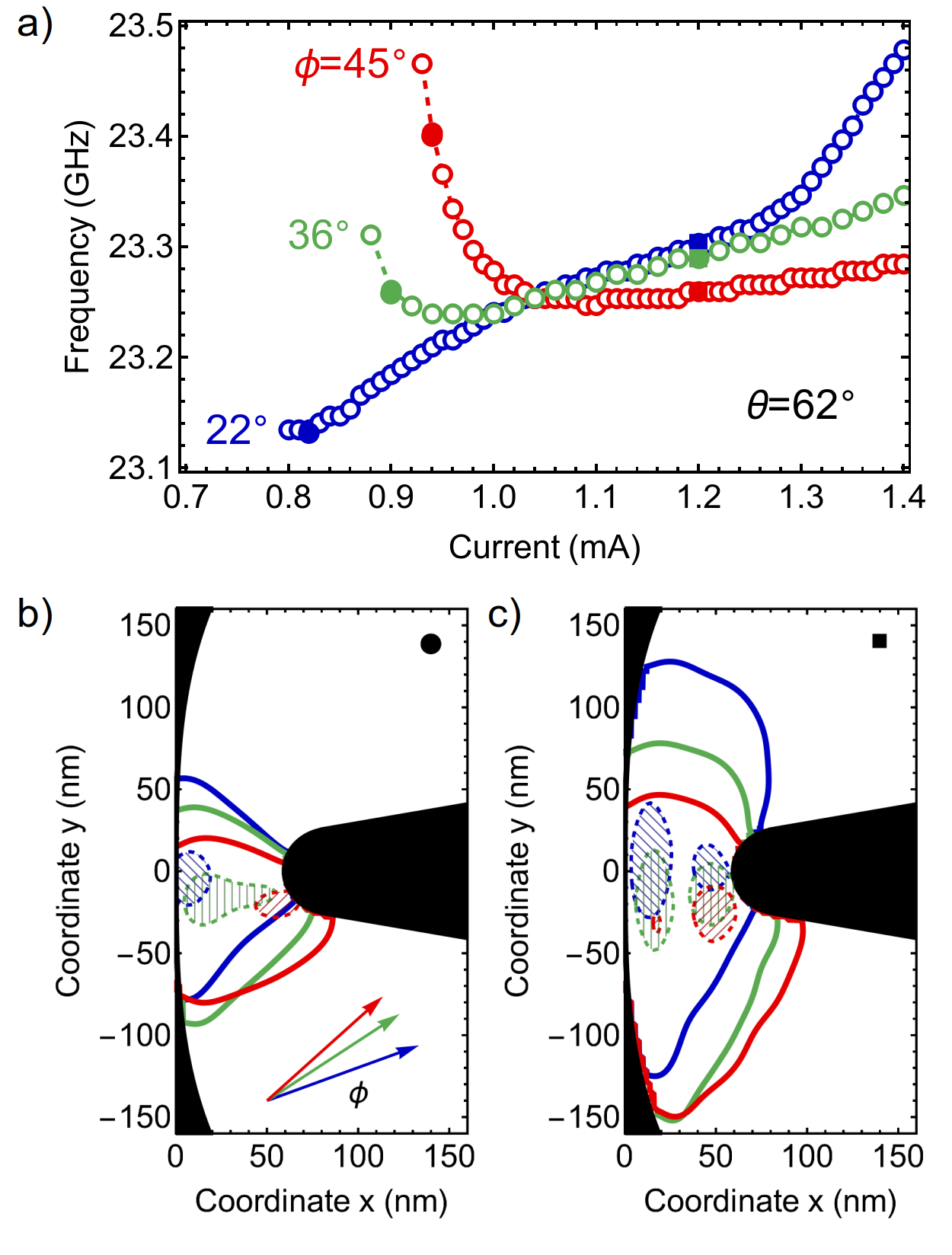}
    \caption{(a) Dependence of the auto-oscillation frequency of the single ANC-SHNO on the applied dc current for several in-plane orientations of the external magnetic field, with the polar angle fixed at $\theta = 62^\circ$. The azimuthal angles are indicated by color: $\phi = 22^\circ$ (blue), $\phi = 36^\circ$ (green), and $\phi = 45^\circ$ (red). (b), (c) Spatial distributions of the auto-oscillation modes, visualized as contours of magnon density corresponding to 50\% (solid lines) and 95\% (dashed lines, hatched regions) of the maximum density. Panel (b) shows the mode profiles just above their respective threshold currents [marked by filled circles in a)], whereas panel (c) presents the profiles at a fixed current of $I = 1.2$~mA [marked by filled squares in a)]. The color coding of the contours corresponds to the field orientations in panel (a).}
    \label{fig:single}
\end{figure}

\bigskip
To establish a baseline for coupling effects, we first characterize the auto-oscillation of an isolated ANC-SHNO. Figure~\ref{fig:single} (a) shows the oscillation frequency versus dc current for several in-plane field orientations. The device exhibits a wide tunability of several hundred MHz, with both the sign and magnitude of the frequency–current slope (nonlinearity) strongly dependent on field azimuth. At $22^\circ$ (blue), the frequency blueshifts monotonically, while at $45^\circ$ (red) it shows an overall negative slope, which slightly increases after $I=1.1$~mA. At an intermediate angle ($\phi = 36^\circ$, green curve), the frequency–current relation displays a clear non-monotonic dependence: the frequency decreases with current at low bias (negative nonlinearity) but increases at higher bias (positive slope) with a turnover point at $I=0.97$~mA.

Figures~\ref{fig:single}(b–c) show that the observed frequency trends correlate with the spatial position of the auto-oscillating mode. At $22^\circ$ and $45^\circ$, the mode remains tightly confined to the outer and inner edges of the constriction, respectively. In contrast, at $36^\circ$, the mode emerges at the threshold within the constriction interior. Moreover, varying the in-plane field angle also shifts the mode center along the constriction. This shift, together with the observed profiles' asymmetry, arises from changes in the relative orientation between the equilibrium magnetization and the spin-current polarization. The curved constriction edge modifies the local charge-current flow, thereby altering spin-current polarization on either side. Deliberately asymmetric edge geometries -- for instance, with enhanced curvature on the outer edge -- could therefore be suitable for field-tunable nonreciprocal devices.

At higher bias, all modes expand in volume. For $36^\circ$ and $45^\circ$, this expansion drives the crossover from negative to positive nonlinearity, as a larger active region stiffens the effective magnetic potential and shifts the oscillation frequency upward at higher power. In the following, we focus on the $36^\circ$ field configuration, which enables the study of mutual synchronization across opposite signs of frequency nonlinearity.

\begin{figure}[hbt!]
    \centering
    \includegraphics[width=\linewidth]{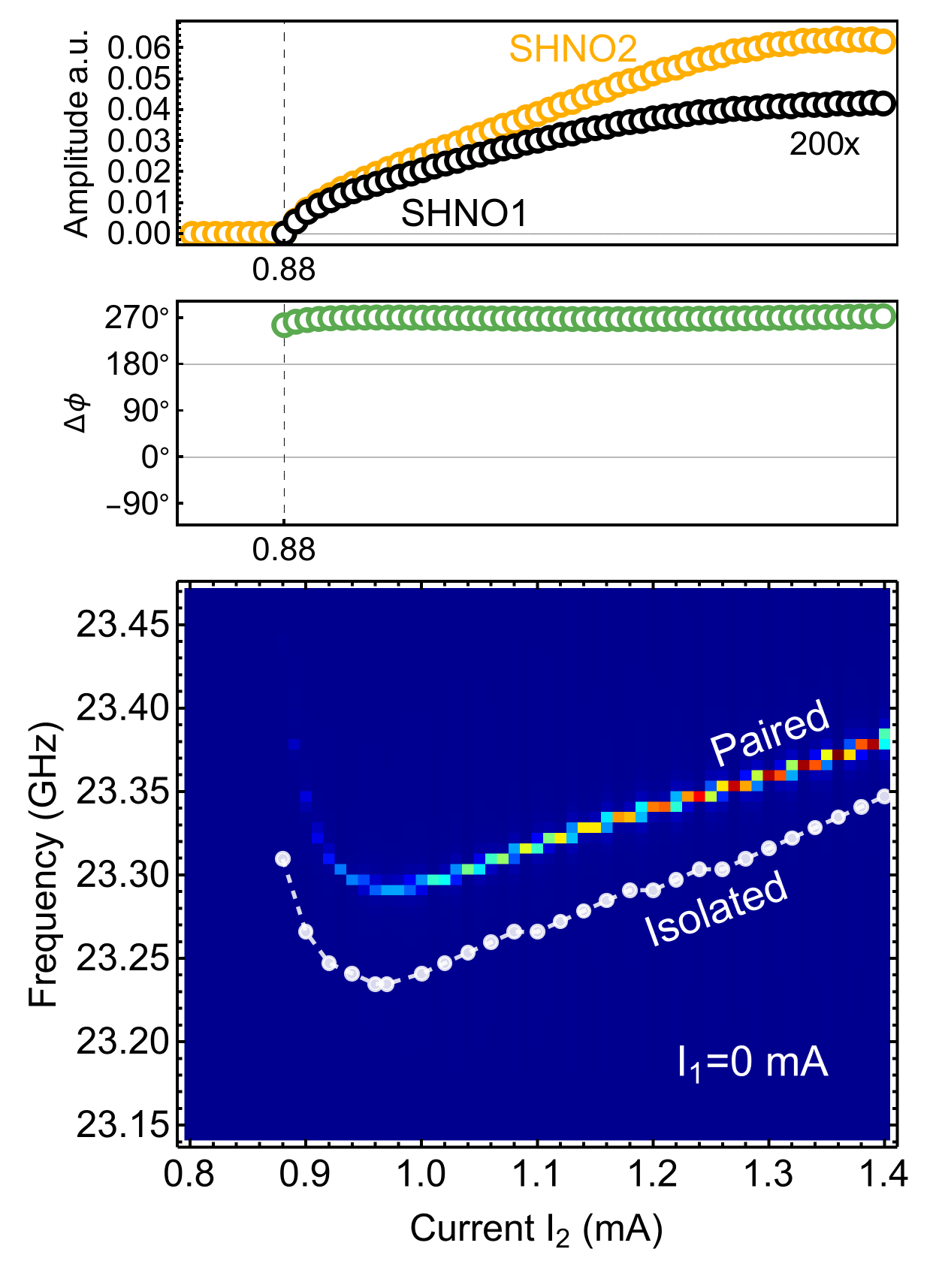}
    \caption{Pair of ANC-SHNOs with SHNO1 turned off, such that it does not self-oscillate but follows SHNO2 passively via dynamic coupling. (Top) Amplitudes of the individual devices at their dominant frequency versus the drive current in SHNO2, $I_2$. For clarity, the SHNO1 amplitude is multiplied by 200. The vertical dashed line marks the auto-oscillation threshold of SHNO2. (Middle) Phase difference $\Delta\phi$ between the two signals at the dominant frequency, showing an approximately constant offset of $-\pi/2$ above threshold. (Bottom) PSD map of the combined signal from both SHNOs as a function of $I_2$. The white curve with dots reproduces the frequency–current dependence of an isolated SHNO from Fig.~\ref{fig:single}a.}
    \label{fig:paired}
\end{figure}

\begin{figure*}[hbt!]
    \centering
    \includegraphics[width=\linewidth]{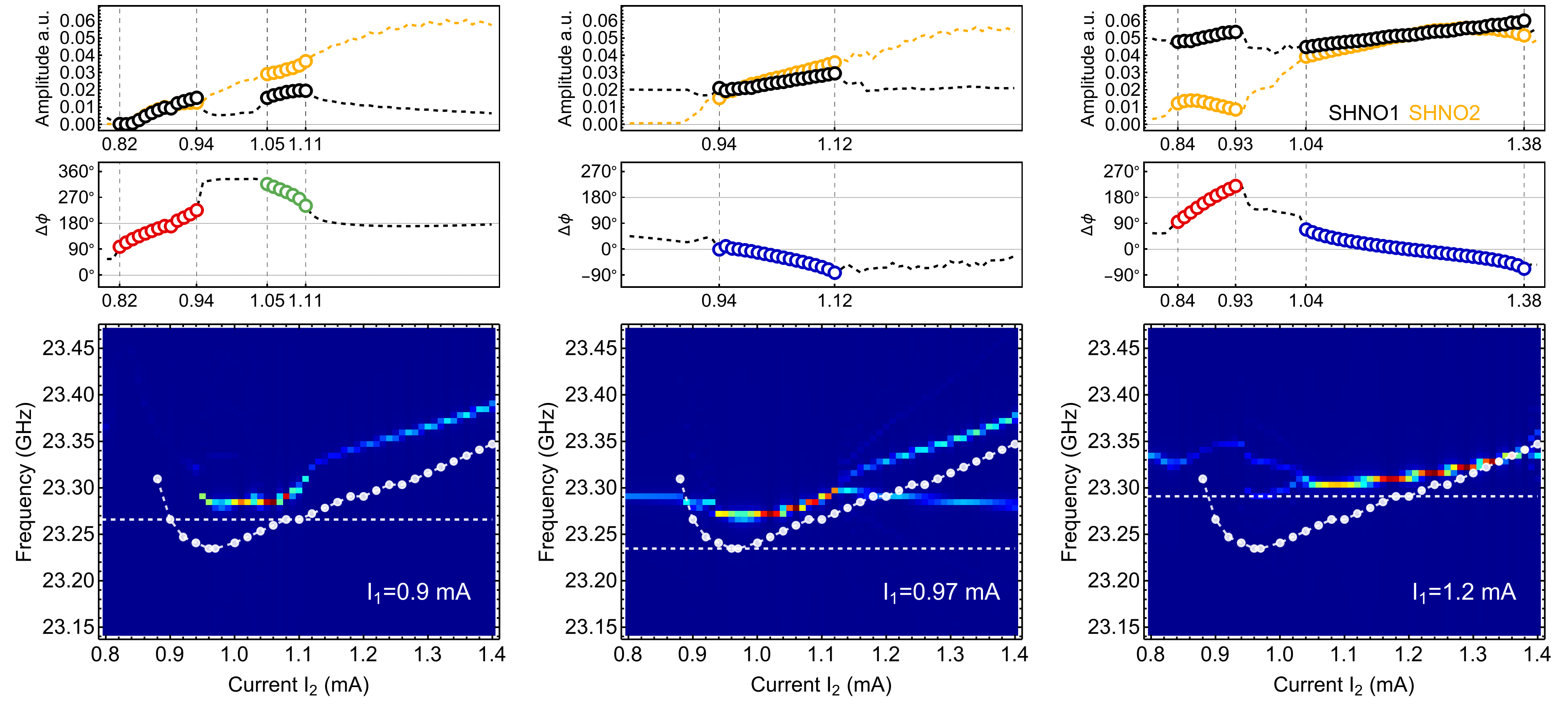}
    \caption{Pair of ANC-SHNOs with finite control currents applied to SHNO1. Each column corresponds to a different fixed $I_1$: (left) $0.9$ mA, (middle) $0.97$ mA, and (right) $1.2$ mA. (Top) Amplitudes of the individual devices at their dominant frequencies versus the drive current in SHNO2, $I_2$. (Middle) Phase difference $\Delta\phi$ between the two signals determined via Fourier analysis. Dashed curves cover the cases when the dominant frequencies in individual spectra may differ. (Bottom) PSD maps of the combined signals as a function of $I_2$. The white dashed curves (with dots) reproduce the frequency–current dependence of an isolated SHNO from Fig.~\ref{fig:single}a.}
    \label{fig:control}
\end{figure*}

\begin{figure*}[hbt!]
    \centering
    \includegraphics[width=\linewidth]{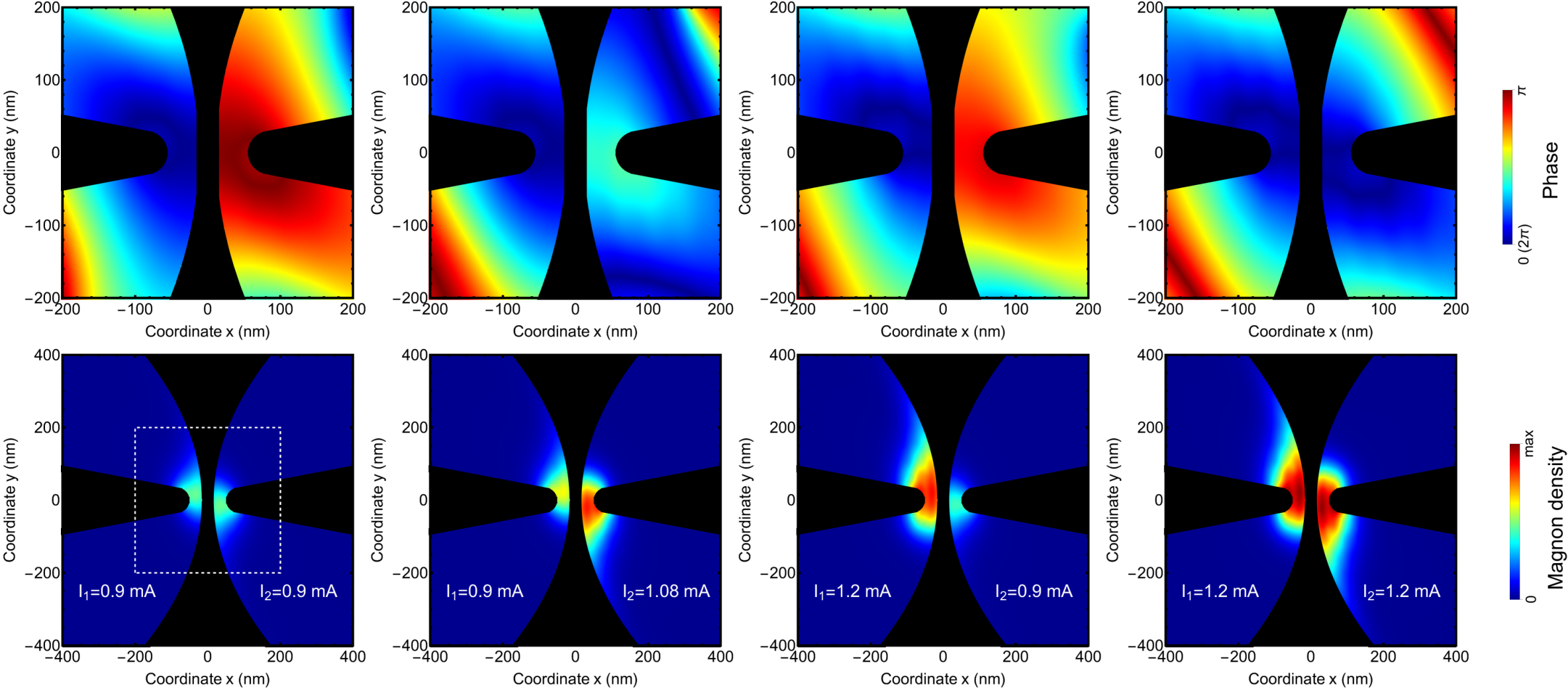}
    \caption{Spatial maps of magnon phase (top row) and normalized magnon density (bottom row) in a pair of ANC-SHNOs under symmetric and asymmetric biasing conditions: (left) $I_1 = 0.9$~mA, $I_2 = 0.9$~mA,  (middle left); $I_1 = 0.9$~mA, $I_2 = 1.08$~mA; (middle right) $I_1 = 1.2$~mA, $I_2 = 0.9$~mA; and (right) $I_1 = 1.2$~mA, $I_2 = 1.2$~mA.}
    \label{fig:maps}
\end{figure*}

To examine a control case for two oscillators, we begin with a unidirectional coupling scenario in which only one oscillator is actively driven. In this setup, SHNO2 is biased with current, while SHNO1 remains off ($I_1 = 0$~mA). Without any coupling, SHNO1 would stay magnetically inactive. However, once SHNO2 begins oscillating, its dynamic stray field induces a response in SHNO1. As shown in Fig.~\ref{fig:paired}, SHNO1 develops a small oscillation (scaled $\times200$ for clarity) that remains far below SHNO2's amplitude, consistent with its lack of direct spin-torque drive. The phase difference $\Delta\phi$ between the two signals equals $-\pi/2$, indicating that SHNO1 lags SHNO2 by a quarter period, characteristic of resonant forcing near the natural frequency. The combined power spectral density (PSD) shown in the bottom panel reveals that the presence of other nearby SHNO results in a frequency shift upward by about 55~MHz compared to the case of the isolated device, preserving its non-monotonic shape.

We next investigate full mutual synchronization of two ANC-SHNOs, with both oscillators biased above threshold. To probe different regimes of amplitudes and nonlinear frequency detuning, $I_1$ is fixed at three representative values - 0.90~mA, 0.97~mA, and 1.20~mA - while $I_2$ is swept in each case. These operating points correspond to SHNO1 being slightly above threshold in the negative frequency–current slope regime (0.90~mA), at the frequency minimum where the nonlinearity changes sign (0.97~mA), and well above threshold in the positive-slope regime (1.20~mA). In all cases, strong mutual coupling is achieved, though the relative phase and amplitude evolution depend sensitively on the bias point of SHNO1. Fig.~\ref{fig:control} details these three biasing scenarios, while Fig.~\ref{fig:maps} connects them to the real-space magnons phase/density snapshots for specific cases.

For a small control current $I_1=0.9$~mA (left column in Fig.~\ref{fig:control}), the combined PSD first displays a noticeable power only once $I_2 \ge 0.96$~mA. Notably, weak auto-oscillations of comparable (small) amplitude actually develop in both oscillators already over $I_2 \in[0.82,0.94]$, see Fig.~\ref{fig:maps} (left) for the case $I_1=I_2=0.9$~mA. In this range of $I_2$, their frequencies are synchronized while the relative phase drifts monotonically and crosses $180^\circ$ near equal currents. Such destructive interference suppresses the net output, causing the signal to vanish or be strongly attenuated in the PSD map. Upon further increasing $I_2$, the dynamics proceeds through three characteristic stages. (i) The natural-frequency detuning exceeds the mutual locking range, so that—while the amplitudes remain similar -- the spectra acquire a comb-like modulation indicative of amplitude/phase beating in the unlocked regime. (ii) As $I_2$ continues to rise, SHNO2 returns to SHNO1's frequency, but now on the positive slope of its nonlinear tuning curve and with a larger amplitude; the phase difference approaches $\sim 270^\circ$, see also Fig.~\ref{fig:maps} (middle left). (iii) Finally, SHNO2's amplitude grows enough to pull SHNO1 onto its branch and enforce out-of-phase ($180^\circ$) locking, which restores a strong line in the combined PSD despite the $\pi$ phase offset. 

When the control current places SHNO1 at the frequency minimum, $I_1 = 0.97$~mA (Fig.~\ref{fig:control}, middle column), the synchronization diagram is canonical. The two frequencies merge at $I_2\approx0.94$~mA and remain mutually locked up to $I_2\approx1.12$~mA. Throughout this interval, the relative phase evolves smoothly, starting from near in-phase, while the amplitudes of both oscillators increase together. Exiting this window, the conservative coupling can no longer compensate for the nonlinear detuning, and the PSD splits into two distinct lines corresponding to the individual oscillators.

For a large control current $I_1 = 1.2$~mA (right column in Fig.~\ref{fig:control}), SHNO1 sustains a strong auto-oscillation and dominates the pair's gain competition. At small $I_2$, the oscillators lock robustly in anti-phase ($180^\circ$), so the net output remains finite due to the amplitude imbalance (i.e., no complete cancellation occurs), see Fig.~\ref{fig:maps} (middle right). In this regime, the common frequency increases with $I_2$, reflecting locking on the positive-slope branch. Interestingly, additional pumping into SHNO2 initially reduces its amplitude -- the stronger oscillator suppresses the weaker through dipolar coupling -- until the system traverses a loss of lock over $I_2 \in[0.93,1.04]$~mA. Once SHNO2 passes its frequency minimum and its amplitude becomes comparable to SHNO1, mutual synchronization is recovered, see also Fig.~\ref{fig:maps} (right). Crucially, with a strong bias at SHNO1, it is possible (by tuning $I_2$) to realize both in-phase ($0^\circ$) and anti-phase ($180^\circ$) states at essentially the same frequency and with non-zero net output, an attribute attractive for phase-binarized arrays and Ising-type computing~\cite{albertsson2021ultrafast, houshang2022phase, litvinenko2023spinwave}.

\bigskip
In conclusion, we introduce a new asymmetric-nanoconstriction design of spin-Hall nano-oscillators and, via micromagnetic simulations, demonstrate mutual synchronization of two independently biased devices mediated solely by dipolar stray fields. To our knowledge, this is the first theoretical demonstration of dipolar-only mutual locking in separately powered SHNOs. Depending on the bias conditions, the pair stabilizes either an in-phase or an out-of-phase state, with the transition governed by the relative amplitudes and nonlinear frequency detuning of the oscillators. Importantly, amplitude imbalance favors the out-of-phase configuration accompanied by partial suppression of the weaker oscillator, whereas balanced operation allows bistability between $0^{\circ}$ and $180^{\circ}$ phase locking. Moreover, the combination of strong local coupling with independent frequency and gain control supports reconfigurable oscillator clusters for neuromorphic computing~\cite{chumak2022advances, finocchio2024roadmap,gonzalez2024spintronic, arunima2025controllable, manna2024phase} and offers a testbed for non-Hermitian spintronic phenomena~\cite{wittrock2024non,perna2024coupling, matveev2023exceptional, flebus2020non}.

\bibliography{main}

\end{document}